\documentclass{article}

\usepackage{microtype}
\usepackage{graphicx}
\usepackage{subfigure}
\usepackage{booktabs} 
\usepackage{fontsize}

\usepackage{hyperref}
\usepackage{listings}
\usepackage{cleveref}
\usepackage{pifont}
\usepackage[dvipsnames]{xcolor}
\usepackage{array}
\usepackage{appendix}
\usepackage{multirow}
\usepackage{multicol}

\usepackage[accepted]{mlsys2025}

\mlsystitlerunning{Declarative Data Pipeline for Large Scale ML Services}

\begin{document}

\twocolumn[
\mlsystitle{Declarative Data Pipeline for Large Scale ML Services}




\author{Yunzhao Yang \quad Runhui Wang$*$ \quad Xuanqing Liu$*$ \quad Adit Krishnan$*$ \quad Yefan Tao$*$ \quad Yuqian Deng$*$ \quad Kuangyou Yao$*$ \quad Peiyuan Sun$*$ \quad  Henrik Johnson \quad Aditi sinha\quad Davor Golac\quad Gerald Friedland\quad Usman Shakeel\quad Daryl Cooke\quad Joe Sullivan\quad Chris Kong  $^\dagger$
  \\ 
$*$ Equal Contribution\\
$^\dagger$ Corresponding Author
\\Amazon Web Services, Seattle, USA
}

\mlsyssetsymbol{equal}{*}

\begin{mlsysauthorlist}

\mlsysauthor{Yunzhao Yang \quad Runhui Wang$*$ \quad Xuanqing Liu$*$ \quad Adit Krishnan$*$ \quad Yefan Tao$*$ \quad Yuqian Deng$*$}{}

\mlsysauthor{\quad Kuangyou Yao$*$ \quad Peiyuan Sun$*$ \quad  Henrik Johnson \quad Aditi sinha\quad Davor Golac\quad Gerald Friedland}{}

\mlsysauthor{\quad Usman Shakeel\quad Daryl Cooke\quad Joe Sullivan\quad Madhusudhanan Chandrasekaran \quad Chris Kong$^\dagger$}{}

\mlsysauthor{Amazon Web Services}{}

\end{mlsysauthorlist}


\mlsyscorrespondingauthor{Chris Kong}{luyankon@amazon.com}

\mlsyskeywords{Machine Learning, MLSys, Large Scale Data Management}

\vskip 0.3in

\begin{abstract}

Modern distributed data processing systems struggle to balance performance, maintainability, and developer productivity when integrating machine learning at scale. These challenges intensify in large collaborative environments due to high communication overhead and coordination complexity. We present a ``Declarative Data Pipeline" (DDP) architecture that addresses these challenges while processing billions of records efficiently. Our modular framework seamlessly integrates machine learning within Apache Spark using logical computation units called Pipes, departing from traditional microservice approaches. By establishing clear component boundaries and standardized interfaces, we achieve modularity and optimization without sacrificing maintainability. Enterprise case studies demonstrate substantial improvements: 50\% better development efficiency, collaboration efforts compressed from weeks to days, 500× scalability improvement, and 10× throughput gains. 
\end{abstract}
]



\printAffiliationsAndNotice{\mlsysEqualContribution} 

\section{Introduction}




Modern big data processing systems face a critical yet often overlooked challenge: the trade-off among system performance, collaborative productivity, system maintainability, and separation of concerns~\cite{mili2004understanding}. While achieving high throughput at massive scale is crucial, equally important is maintaining code that developers can efficiently understand, collaborate, modify, and test. 
Distributed computing frameworks like Apache Spark excel at processing petabyte-scale datasets~\cite{shanahan2015large}, but the optimizations required for peak performance typically result in tightly coupled, hard-to-maintain codebase. Furthermore, as data pipelines grow in complexity, developers face an increasingly daunting cognitive load, making it difficult to confidently reason about system behavior or troubleshoot and implement fixes fast when large-scale issues arise.

Our analysis of existing solutions reveals fundamental limitations in addressing this optimization-productivity-maintainability tradeoff. Traditional microservice-based architectures struggle with large-scale data processing due to excessive network communication overhead. While SystemML\cite{boehm2016systemml} is powerful, its cost-based optimization leads to unpredictable execution plans with dynamic data. Cedar\cite{zhao2024cedar} addresses this through composable operators, but its reliance on profiling and optimization heuristics makes it vulnerable to variable data patterns.

To address these challenges, we introduce a novel Declarative Data Pipeline (DDP) architecture built upon the \textit{``pipe"} - a modular component with well-defined inputs and outputs that perform a specific data transformation. 
Similarly to a microservice, one pipe functions as a standalone computational unit. The difference is that instead of communicating over network among microservices, these pipes are chained together via system memory for high throughput. This approach benefits from the modularity and isolation like microservices while avoiding the network overhead typically associated with REST API calls. Our architecture's logical separation between pipe implementations and the input-output signature of each pipe provide several key advantages over existing solutions. By establishing clear workflow boundaries for each pipe, we enable fully parallel development and isolated logic testing. We provide end-to-end pipeline workflow visualization for integration test which dramatically simplify testing compared to systems like Cedar~\cite{zhao2024cedar}. Through standardized interfaces, we achieve different levels of granularity of code and data reusability beyond what current workflow-as-code solutions offer. Lastly, our declarative pipeline definitions improve system maintainability while preserving the performance benefits demonstrated by solutions like SystemML~\cite{boehm2016systemml}. We note that our approach applies to any data process, not just for ML workflows.


\begin{figure*}[h]
    \centering
    \includegraphics[width=15cm]{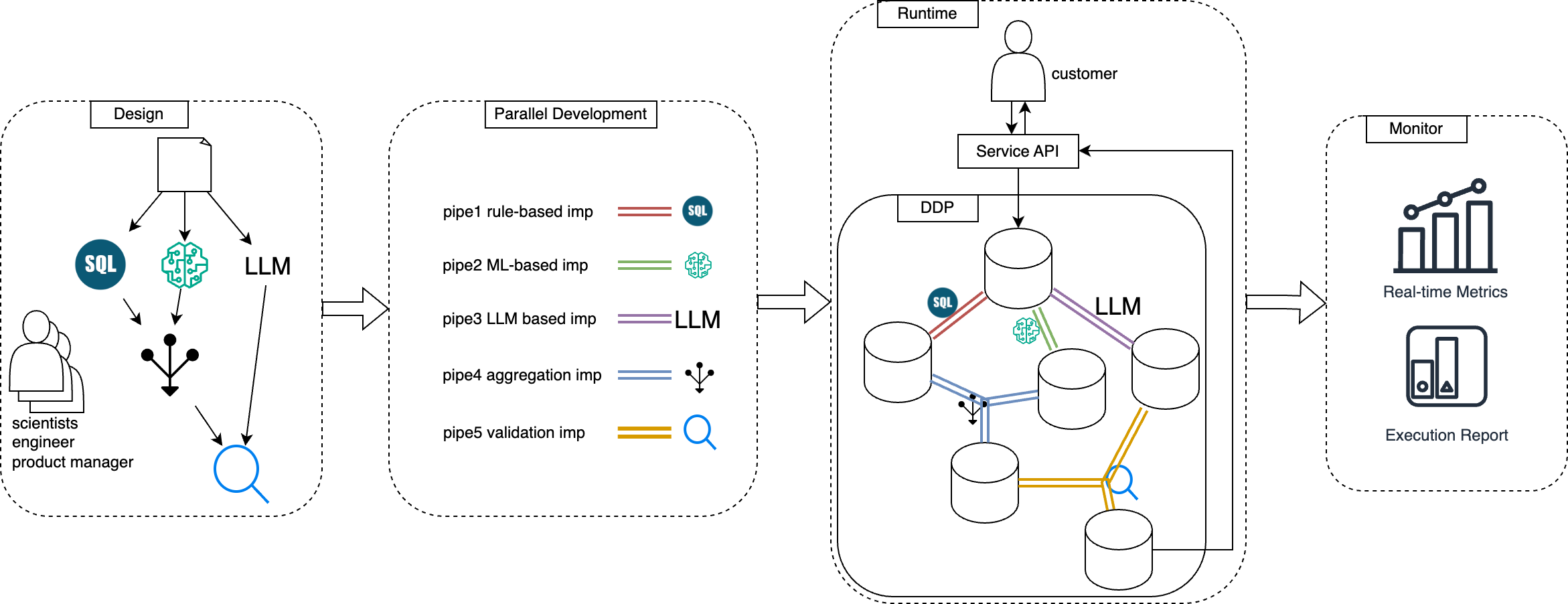}
    \caption{This represents the full development cycle using DDP for a product leveraging SQL, traditional model and LLM, from design, development, runtime to monitoring.}
    \label{DDP}
\end{figure*}

Our evaluation of Declarative Data Pipeline demonstrates its exceptional performance across both industry and research applications. Figure \ref{DDP} shows the full development cycle using DDP to develop an example large scale ML system leveraging SQL rules, traditional model and LLM. In the enterprise case study (Table \ref{industry}), DDP delivered transformative efficiency gains: development cycles shortened by 50\%, codebase size reduced by 40\%, and integration/troubleshooting efforts compressed from weeks to days. Performance metrics were equally impressive, with 500$\times$ improved scalability and 10× higher throughput. The throughput improvements were reinforced by our academic language detection experiment (Table \ref{web}), where DDP outperformed non-distributed implementations by 180× and surpassed Ray-based solutions by 5.7$\times$. Beyond raw performance, DDP significantly enhanced system maintainability through built-in real-time metrics and workflow visualization-capabilities that required minimal additional code. These comprehensive improvements demonstrate that DDP successfully resolves the longstanding trade-off between system optimization and developer productivity, delivering exceptional results on both dimensions simultaneously.

From the engineering perspective, we made several key architectural decisions to optimize both the system performance and development efficiency and velocity. The following are the most important architectural considerations. 

\vspace{-0.5em}
\subsection*{Choice of Processing Framework}
We selected Apache Spark as our core processing framework as it is the most widely adopted scalable computing (adopted by thousands of companies, including 80\% of the Fortune 500). Spark offers superior processing efficiency for handling of joins, maps, and reduces, built-in fault tolerance, and effective load balancing, critical features for large-scale data processing systems. When evaluating cloud-based implementations, AWS Glue and Amazon EMR emerged as the most suitable Spark-based services, significantly outperforming alternative technologies such as AWS Lambda, Amazon ECS, or AWS Batch for high-throughput data processing.


\vspace{-0.5em}
\subsection*{ML Model Integration Strategy}
A critical innovation in our architecture is the embedding of ML models directly within the Spark cluster, departing from traditional microservice-based approaches. Conventional wisdom suggests deploying ML models (typically developed in Python) as separate microservices accessed via REST APIs. When deploying ML models as microservices, REST API calls introduce significant overhead, with network latency ranging from 20-100ms per call based on analysis \footnote{https://www.nickolinger.com/blog/2022-12-05-performance-toolbox-measuring/}. Combined with the standard deep learning module such as 12-layer BERT~\cite{liu2019roberta} encoder which takes up to 5ms per record at inference time, this makes a notable performance impact. Our benchmarks revealed that this approach introduces significant performance penalties, with microservice-based integration showing up to 10x lower throughput compared to our embedded approach.

\vspace{-0.5em}
\subsection*{Implementation Language Selection}
We choose Scala as our primary implementation language driven by both performance and architectural considerations. While Python is the de facto standard for ML model development, particularly with frameworks like BERT, using PySpark introduces substantial overhead through data serialization/deserialization between Python and the JVM in production system based on the benchmark \footnote{https://medium.com/@danniesim/faster-and-more-concise-than-udf-spark-functions-and-higher-order-functions-with-pyspark-31d31de5fed8}.

To bridge this gap, we developed a novel approach to compile Python ML models into Java using Open Neural Network Exchange (ONNX\footnote{https://onnx.ai/}), enabling direct embedding within our Scala-based pipeline. This technique maintains the benefits of Python for ML development while achieving native JVM performance in production. (saving back and force seralization overhead between python and Spark running in JVM which can be up to 10x of the processing latency)

These architectural decisions collectively enable our pipe-based architecture to achieve both high performance and improved developer productivity. The following sections provide a detailed analysis of our design choices, implementation strategies, and performance results in production environments.

\subsection*{Main Contributions}
We highlight that the core contribution of our work is the declarative, memory-bound pipe abstraction that:
\begin{itemize}
    \item replaces network-bound microservices with in-memory contract-driven modules, eliminating REST overhead of 20 to 100 ms per call;
    \item does not rely on cost-based optimization but on deterministic DAG execution driven by declarative definitions, unlike other frameworks like Cedar or SystemML;
    \item unifies declarative configuration, in-memory ML execution, and Spark-based data orchestration in a single execution layer.
\end{itemize}

\section{Related Work}


\subsection{ML workflow design}

Recent research has addressed ML integration in distributed systems. SystemML\cite{boehm2016systemml} provides declarative ML with automatic optimization across computing environments but faces memory constraints in complex scenarios. Potla et al.\cite{potla2022scalable} improve ML pipeline scaling through data partitioning and resource management, though communication overhead persists. Cedar~\cite{zhao2024cedar} offers composable operators with optimizations like caching and fusion, but struggles with unpredictable data patterns. TFX is an end-to-end platform for deploying production ML pipelines \footnote{https://www.tensorflow.org/tfx}. Delta live tables \footnote{https://www.databricks.com/blog/2021/05/27/announcing-the-launch-of-delta-live-tables-reliable-data-engineering-made-easy.html} simplifies ETL development and management for DataBricks customers.

Gap: While these systems optimize distributed ML execution, they lack integrated visual development tools and unified support for both local debugging and production deployment, limiting developer productivity in iterative workflow design.

\vspace{-1em}
\subsection{Workflow as code}

Modern orchestration systems like Apache Airflow~\cite{haines2022workflow}, Prefect~\cite{narayanan2024orchestrating}, Kubeflow Pipelines~\cite{bisong2019kubeflow}, and Flyte~\cite{flyte} enable programmatic DAG-based workflow definition with version control. However, they face debugging and maintenance challenges~\cite{singla2023machine}, with usability and visualization issues~\cite{ono2020pipelineprofiler}. Recent work~\cite{zhou2023instructpipe, heffetz2020deepline, nikitin2022automated} shows visual approaches improve efficiency.

Gap: Existing workflow systems either prioritize code-first approaches with poor visualization or provide visual tools without seamless Spark integration and local-to-production workflow portability.

\subsection{Comparison with other framework}

\newcolumntype{P}[1]{>{\centering\arraybackslash}p{#1}}
\newcommand{\cmark}{\textcolor{ForestGreen}{\ding{51}}} 
\newcommand{\xmark}{\textcolor{BrickRed}{\ding{55}}}  

\begin{table*}[hbt!]
    \small
    \caption{Comparison of Framework Capabilities: Core Features}
    \centering
    \begin{tabular}{|c|P{1.6cm}|P{1.3cm}|P{1.8cm}|P{1.6cm}|P{2cm}|}
    \hline
    Framework & Distributed Computing & Big Data Support & Spark Runtime Integration & Spark Dev Integration & Spark Dev Method \\
    \hline
    DDP & \cmark & \cmark & \cmark & \cmark & JAR \\
    \hline
    SystemML \cite{boehm2016systemml} & \cmark & \cmark & \cmark & \cmark & JAR/Notebook \\
    \hline
    Cedar~\cite{zhao2024cedar} & \cmark & \cmark & \xmark & \xmark & \\
    \hline
    Flyte\cite{haines2022workflow} & \cmark & \cmark & \cmark & \cmark & JAR/Notebook \\
    \hline
    Ray\cite{moritz2018ray} & \cmark & \cmark & \cmark & \cmark & JAR/Notebook \\
    \hline
    AWS Step Function\cite{aws-step-function} & \cmark & \cmark & \cmark & \xmark &  \\
    \hline
    DataBricks\cite{databricks} & \cmark & \cmark & \cmark & \cmark & Notebook \\
    \hline
    AWS Glue\cite{saxena2023story} & \cmark & \cmark & \cmark & \cmark & JAR/Notebook \\
    \hline
    AWS EMR\cite{aws-emr} & \cmark & \cmark & \cmark & \cmark & JAR/Notebook \\
    \hline
    AWS ECS / Batch\cite{aws-ecs}& \cmark & \xmark & \xmark & \cmark &  \\
    \hline
    AWS Lambda\cite{aws-lambda} & \cmark & \xmark & \xmark & \xmark &  \\
    \hline
    Native Spark\cite{zaharia2016apache} & \cmark & \cmark & \cmark & \cmark & JAR/Notebook \\
    \hline
    \end{tabular}
    
    \label{tab:frameworks_part1}
\end{table*}

\begin{table*}[hbt!]
    \small
    \centering
    \caption{Comparison of Framework Capabilities: Workflow \& Interface Features}
    \begin{tabular}{|c|P{2cm}|P{2cm}|P{1.8cm}|P{2.3cm}|}
    \hline
    Framework & Multi Step Workflow & Cluster Management & UI Assistant & Spark Interface \\
    \hline
    DDP & \cmark & \xmark & \cmark & \cmark \\
    \hline
    SystemML~\cite{boehm2016systemml} & \cmark & \cmark & \xmark & \cmark \\
    \hline
    Cedar~\cite{zhao2024cedar} & \xmark & \cmark & \cmark & \xmark\\
    \hline
    Flyte~\cite{haines2022workflow} & \cmark & \cmark & \xmark & \cmark \\
    \hline
    Ray~\cite{moritz2018ray} & \cmark & \cmark & \xmark & \xmark \\
    \hline
    Airflow~\cite{haines2022workflow} & \cmark & \xmark & \xmark & \xmark \\
    \hline
    AWS Step Function\cite{aws-step-function} & \cmark & \xmark & \xmark & \xmark \\
    \hline
    DataBricks~\cite{databricks} & \cmark & \cmark & \xmark & \cmark \\
    \hline
    AWS Glue~\cite{saxena2023story} & \cmark & \cmark & \xmark & \cmark \\
    \hline
    AWS EMR~\cite{aws-emr} & \xmark & \cmark & \xmark & \cmark \\
    \hline
    AWS ECS / Batch~\cite{aws-ecs}& \xmark &  & \xmark & \cmark \\
    \hline
    AWS Lambda~\cite{aws-lambda} & \xmark &  & \xmark &  \\
    \hline
    Native Spark\cite{zaharia2016apache} & \xmark & \xmark & \xmark & \cmark \\
    \hline
    \end{tabular}
    
    \label{tab:frameworks_part2}
\end{table*}

We provide a qualitative evaluation comparing the declarative data pipeline with other popular ML big data frameworks in the following dimensions in Table \ref{tab:frameworks_part1} and \ref{tab:frameworks_part2}. 

\begin{enumerate}
    \item \textbf{Distributed Computation}: Support for running  in an horizontal scalable environment
    \item \textbf{Big Data Support}: Provide integration with established data storage or streaming platform (e.g: AWS S3 / kinesis)
    \item \textbf{ML Integration}: Support integration with popular ML framework (Pytorch, DJL, etc)
\item \textbf{Spark Development Integration}: Developer can create local executable workflows for debugging and tests
\item \textbf{Spark Development Method}: How could developer run spark jobs
\item \textbf{Multi step workflow}: Support multi steps workflow orchestration (e.g: A sequence of Tokenization $\rightarrow$ Embedding $\rightarrow$ KNN clustering workflow)
\item \textbf{Cluster Management}: Ability to manage clusters like YARN
\item \textbf{UI assistant}: Provides UI to support visualization of workflow and template code generation
\item \textbf{Spark Interface}: Ability to directly control spark runtime configuration (e.g: executor heartbeat timeout)
\end{enumerate}

\section{Methodology}

This section presents the Declarative Data Pipeline framework's methodology. We introduce core architecture components, including Data as Anchor, Logic Unit as Pipe, and Explicit State Management. The framework's system features and implementation details are then discussed, covering Out-of-box Functionalities,  Data-Driven Execution Flow, Data Flow Control, and Object Life-cycle Optimization. 
Throughout this section, we demonstrate how these elements collectively support both high-performance processing of large-scale datasets and improved developer productivity in ML-integrated distributed environments, particularly when handling billions of records with complex transformation requirements.

\subsection{Data as Anchor, Logic unit as Pipe}
\label{data_as_anchor}

The fundamental goal of a data pipeline is to transform raw input into desired output through a series of intermediate transformations. While the transformation logic is important, we prioritize the data itself - the inputs, outputs, and intermediate data - as the primary focus of our design. We implement this using declarative programming, where all dataset properties (including location, schema, file type, and encryption settings) are explicitly defined at the program entry point. 
By treating these intermediate datasets as ``\textbf{anchor}", we naturally decompose the pipeline system into well-defined modular components called ``\textbf{pipes}". A pipe, as a collection of logic using Spark RDD, DataSet or Dataframe, represents a well-defined transformation unit that consumes specific inputs and produces the defined output. To ensure system-wide compatibility and ease of integration, we introduce a standardized interface that allows these Pipes to be seamlessly connected into complete pipelines. The fundamental structure of a pipe follows a simple paradigm:

{\footnotesize
\begin{verbatim}
Inputs → Pipe (Transformation Logic) → Outputs
\end{verbatim}
}
This modular design offers three key advantages. First, it enables engineers to focus exclusively on implementing core computation logic, as peripheral concerns such as data I/O, encryption, metrics tracking, and execution orchestration are handled automatically by the infrastructure. Second, it promotes high-level modularity and reusability, allowing complex systems to be decomposed into chainable components. Third, by maintaining clear input-output contracts, each pipe encapsulates specific functionality, enabling independent development and testing. This standardization significantly reduces integration complexity and promotes code reusability across different pipeline implementations. Below is an example of an ML data pipeline definition with preprocessing, feature generalization, model prediction, and post-processing steps: \\

\vspace{-1em}

{\scriptsize
\begin{verbatim}
{"inputDataId": ["InputData"],
 "transformerType": "PreprocessTransformer",
 "outputDataId": "IntermediateData"
},
{"inputDataId": ["InputData"],
 "transformerType": "FeatureGenerationTransformer",
 "outputDataId": "FeatureData"
},
{"inputDataId": ["IntermediateData", "FeatureData"],
 "transformerType": "ModelPredictionTransformer",
 "outputDataId": "PredictionData"
},
{"inputDataId": ["InputData", "PredictionData"],
 "transformerType": "PostProcessTransformer",
 "outputDataId": "OutputData"
}
\end{verbatim}}

This architecture provides clear governance over all datasets being consumed and generated, while establishing transparent data lineage for monitoring purposes. Furthermore, this data-centric approach enables parallel development of pipe components across teams. Figure \ref{pipe_illustration} demonstrates this approach with nine dataset declarations:

\begin{figure}[h]
    \centering
    \includegraphics[width=8.4cm]
    {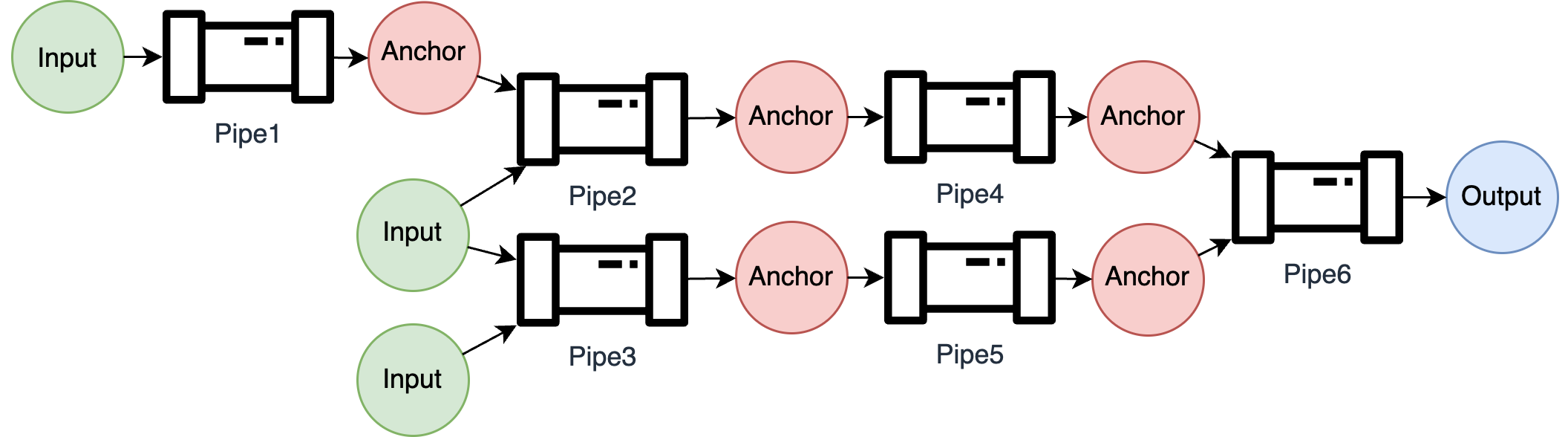}

    \caption{Dataset declarations serve as ``anchor" in our pipeline architecture, specifying data attributes like location, schema, and encryption settings. These declarations form interfaces between pipe components, enabling modular development and independent data processing units.}
    \label{pipe_illustration}
    
\end{figure}

\subsection{Explicit State Management}
\label{explicit_state_management}

While Apache Spark core processing is inherently stateless, our framework implements careful and explicit state management to ensure system reliability and performance. Using the analogy of a water system, we maintain gauges to track aggregated metrics, while avoiding the accumulation of data within the processing pipeline. Our state management strategy encompasses three key aspects:\\
First, we maintain a predominantly stateless architecture to handle unbounded input sizes safely. While this is our default approach, we selectively cache intermediate results to optimize performance. For example, in a data lineage A→B→C→D, reprocessing C would typically require recomputing the entire chain A→B→C. By strategically persisting node C, we can avoid redundant computation when C is required by multiple downstream processes (e.g., C→D and C→E), while avoiding the overhead of storing the complete lineage. Second, we implement the explicit state management within each pipe component, like the "delete" clause in C++. This includes a built-in cleanup mechanism through which the developer can register the internally cached data for removal upon completion of this pipe, preventing resource leaks, and ensuring efficient memory utilization. Third, we incorporate a metrics collection system that enables real-time monitoring without compromising the pipeline's stateless nature. These metrics are automatically published to monitoring systems (e.g., AWS CloudWatch) at configurable cadence (30 seconds by default), providing visibility into the pipeline's performance and health without requiring persistent all states within the processing components.

\subsection{Out-of-box features within Pipe}
\label{out_of_box_feature}
Our architecture implements a modular design that encapsulates distinct logic units within individual pipes. Following object-oriented principles, each pipe is instantiated as a class implementing a generic interface, while adhering to functional programming paradigms for the core transformation function. This design pattern allows us to abstract away common complexities through several out-of-box features, allowing people to focus only on the implementation of the transformation functionality that handles the inputs from memory and generates the output in memory.

\subsubsection{Data I/O Abstraction}
The framework provides unified data access interfaces that support multiple storage systems and file formats. This abstraction layer handles read/write operations across various backends (including distributed file systems, local storage, and NoSQL databases) and supports multiple data formats (CSV, JSON, Parquet, etc.). This separation of concerns allows developers to focus on transformation logic without handling data access and persistence complexities.

\subsubsection{Testing and Debugging Infrastructure}
We implement comprehensive testing capabilities that enable independent validation of each pipe component without requiring external dependencies or cloud service deployment. The architecture supports both unit testing of individual components and integration testing of complete pipelines in local environments, significantly reducing the development-deployment time and improving code readability.

\subsubsection{Security Integration}
The framework incorporates a sophisticated encryption management system that supports multiple security models: service-side encryption/decryption - encrypting all datasets using the same encryption key, dataset-level client-side encryption - encrypting different datasets using different keys, and record-level client-side decryption - encrypting different records using different keys. These security features are configured declaratively through the data specification and handled by the infrastructure, which is separate from the core transformation logic implementation. 

\subsubsection{Metrics and Monitoring}
We implement an asynchronous metrics collection system that provides near real-time visibility into pipeline performance. The framework automatically aggregates and publishes metrics at a configurable cadence (30 seconds by default) without requiring explicit handling within individual pipe components. These metrics were defined individually in each pipe for monitoring different scenarios, enabling comprehensive monitoring while maintaining a clean separation of concerns.

\subsubsection{Platform Independence}
While our implementation mainly targets Apache Spark, the architecture supports platform-agnostic execution through an adapter. This design enables pipes to run across various environments (distributed clusters, single nodes, or cloud services) without modification. We achieve this through a context abstraction layer that standardizes platform-specific interactions. This cross-platform flexibility eliminates code rewrites when moving between Spark environments - whether EMR, Glue, or local setups - saving significant engineering effort, while giving organizations the freedom to choose the most suitable infrastructure for their evolving needs. Furthermore, the pipe interface design extends beyond Spark-specific implementations, allowing integration with non-distributed computing platforms. This flexibility enables the creation of hybrid pipelines that combine distributed and non-distributed processing components while maintaining consistent interfaces and execution patterns.

\subsection{Dynamic Pipe Integration}
\label{dynamic_pipe_integration}
Our framework implements a flexible plugin architecture that facilitates the seamless integration of new pipes through dynamic discovery mechanisms. This approach enables modular system expansion while maintaining architectural consistency. The system employs a runtime discovery mechanism similar to modern dependency injection frameworks, allowing pipes to be dynamically loaded based on configuration specifications. This dynamic loading capability supports flexible pipeline composition without requiring modifications to the core framework. The pipeline structure itself is defined through a declarative configuration format, which specifies the pipe relationships and data dependencies. This plugin-based architecture offers several advantages: it enables independent development and deployment of pipes, supports runtime reconfiguration of processing workflows, and maintains clear separation between pipe implementation and system integration. The declarative nature of the configuration also improves system maintainability by providing a clear, human-readable representation of the pipeline structure and pipe relationships.

\subsection{Data-Driven Execution Flow}

The framework adopts a data-driven approach to workflow management, where the control flow emerges naturally from data dependencies between pipes. Rather than explicitly programming execution sequences, we first generate the data DAG based on the declared input/output relationship that one pipe's input is the upstream of the pipe's output, and then derive the pipe execution order from the data DAG. This approach significantly simplifies pipeline development and maintenance while reducing the potential for control flow errors. The execution order is determined through topological sorting of the data dependency graph, with built-in cycle detection to prevent deadlocks. Each pipe's execution is triggered automatically when its input dependencies are satisfied, creating a self-organizing workflow that adapts to the natural flow of data through the system.

\subsection{Pipeline Visualization}

To support system monitoring and debugging, the framework implements comprehensive visualization capabilities based on the analyzed data DAG that render the whole data pipeline and execution metrics in real-time. We leverage the GraphViz\footnote{https://graphviz.org/} library to render the visualization in real-time. These visualizations provide insights into pipeline structure, execution order, data location, execution progress, and performance characteristics, enabling efficient system optimization and troubleshooting. The combination of automated execution ordering and visual monitoring creates a robust framework for managing complex data processing workflows while maintaining system transparency and operational visibility. Figure \ref{progress_viz} presents the in-progress workflow visualization for the previous example ML data pipeline:

\begin{figure}[h]
    \vspace{1em}
    \centering
    \includegraphics[width=8.2cm]
    {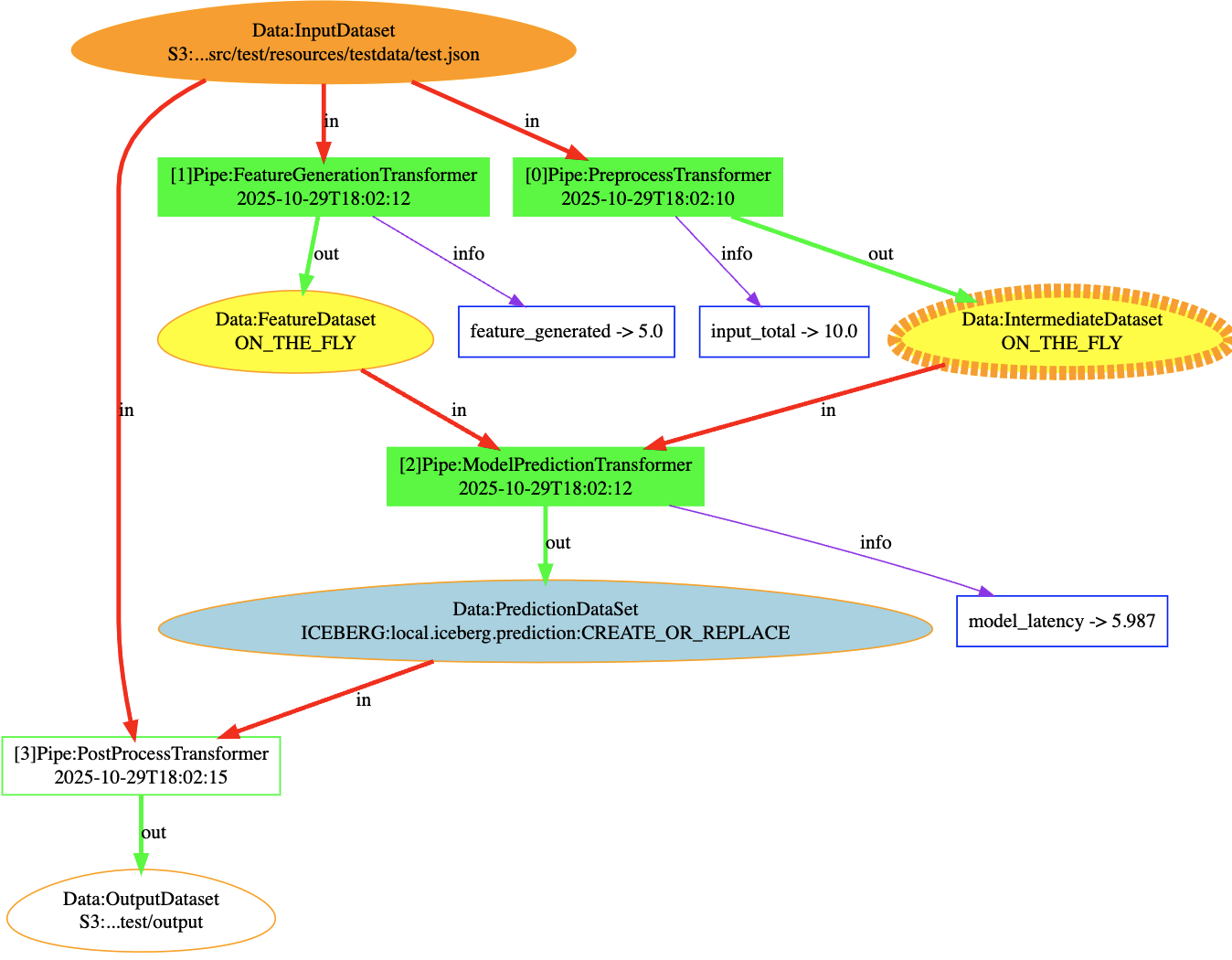}
    \caption{Workflow visualization for an ML data pipeline definition with preprocessing, feature generalization, and model prediction steps. We add the pipe execution order as the prefix to each pipe name such as [0] for the first one. We use purple block with info tag to show the metrics information for each pipe, such as the model\_latency for the ModelPredictionTransformer pipe. We use different colors to notate data with different locations: orange for AWS S3, yellow for memory, dotted orange outline for caching in memory, blue for Iceberg table. We also notate the execution progress with different stages: green for completed steps, yellow for in-progress steps, and white for steps not started.}
    \label{progress_viz}
\end{figure}

\subsection{Object Lifecycle Optimization}
\label{object_lifecycle_optimization}
In distributed computing environments, object initialization and lifecycle management present significant challenges, particularly within pipe implementations using object-oriented programming paradigms. The distributed nature of computation, combined with lazy evaluation strategies, requires careful consideration of object instantiation patterns. The framework identifies and supports three distinct lifecycle scopes for object initialization:

\begin{itemize}
    \item \textbf{record-level}: objects initialized for each data record, leading to multiple instantiations across the data processing cycle.
    \item \textbf{partition-level}: objects created once per data partition, reducing the initialization overhead compared to record-level.
    \item \textbf{instance-level}: each instance only initialized objects once as singleton, providing optimal resource utilization across multiple datasets and processing cycles.
\end{itemize}
The implementation prioritizes instance-level scope optimization, particularly for computationally expensive operations like machine learning model loading and client resource management. This approach significantly reduces initialization overhead and resource consumption compared to conventional per-record or per-partition initialization strategies. By carefully managing object lifecycles at the instance level, the framework achieves efficient resource utilization while maintaining the benefits of distributed processing capabilities. This optimization is especially crucial for resource-intensive objects, such as machine learning models requiring substantial initialization data.

\vspace{-0.5em}
\subsection{Self-service Ecosystem}
\label{self_service_ecosystem}

The framework's modular architecture enables a self-service ecosystem where pipe components can be freely composed and reused. Through standardized interfaces and the anchor-based configuration approach discussed earlier, the system supports dynamic composition of data processing workflows from a repository of pre-validated pipes. Each pipe maintains a strict contract for input and output specifications, enabling automatic validation of pipeline configurations and ensuring compatibility between connected components. The ecosystem approach facilitates rapid pipeline development through component reuse and composition, where users can select and configure qualified pipes from a centralized pipe repository while the framework automatically validates connection compatibility and generates executable workflows. This self-service model significantly reduces development overhead and enables efficient creation of new data processing pipelines without requiring extensive technical expertise or system-level modifications, while the framework's built-in validation ensures that only compatible pipes can be connected, maintaining system integrity while providing flexibility in pipeline design.

\section{Experiments}

In this section, we present one enterprise case study for large-scale data processing, and one academic experiment as language detection for web-scale data designed to evaluate the effectiveness of our DDP framework. We compare the performance of implementations with and without our framework in three key dimensions: development efficiency, runtime performance, and system maintainability.

\vspace{-0.5em}
\subsection{Metrics Introduction}
To evaluate the effectiveness of the DPP framework, we propose a set of quantifiable metrics designed to capture key aspects of software development through three critical dimensions: development efficiency as D, runtime performance as R, and system maintainability as M. Development efficiency measures aspects of development effort and collaboration. Runtime performance assesses processing latency and scalability. System maintainability considers the monitoring effort and troubleshooting effort. These metrics provide quantifiable assessment from both technical and operational perspectives, delivering valuable insights into the framework's immediate development benefits as well as its sustainable value in production environments.

\subsubsection{Development Efficiency} 

\vspace{-0.5em}
\begin{itemize}
\item Development Effort: 
\vspace{-0.5em}
        \begin{itemize}
        \item Feature Development Effort: the number of computation units and lines of code required for the development
        \end{itemize}

\item Collaboration:
\vspace{-0.5em}
        \begin{itemize}
        \item Task Development Parallelism: the number of concurrent tasks and task integration effort measured in days
        \end{itemize}

\end{itemize}

\vspace{-0.5em}
\subsubsection{Runtime Performance} 

\begin{itemize}
\item Computation Efficiency: 
\vspace{-0.5em}
        \begin{itemize}
        \item Execution Time: The total time for the execution
        \item Scalability: The max amount of data can be processed
        \item CPU Utilization Rate: The percentage of available processing capacity being actively used

        \end{itemize}

\end{itemize}

\vspace{-0.5em}
\subsubsection{System Maintainability} 

\begin{itemize}
\item Monitoring Effort:
\vspace{-0.5em}
        \begin{itemize}
        \item Troubleshooting Effort: the development effort to troubleshoot a problem measured in days
        \end{itemize}

\end{itemize}

\subsection{Industry Large-Scale Batch Processing Project}
The DDP approach addresses a critical challenge in industrial big data projects: enabling effective collaboration among large teams working with numerous interdependent computation units. Unlike academic experiments that focus on proof-of-concept implementations, DDP demonstrates measurable productivity improvements in complex production environments.

A real-world case study from a major cloud provider offers compelling evidence of these benefits. The key metrics are shown in Table~\ref{industry}. The project involved processing data at billion-record scale with a team of more than 30 developers collaborating simultaneously. The initial implementation using native Spark encountered significant performance bottlenecks that threatened the entire project timeline. After redesigning the system using DDP, the team successfully eliminated all performance constraints while maintaining the logical separation of components—demonstrating DDP's ability to balance engineering modularity with computational efficiency at enterprise scale.

DDP enabled the team to decompose our solution into ten well-defined, independent pipes, allowing developers to work concurrently without dependencies. The architecture's cross-platform design eliminated cloud deployment requirements during development, as each pipe could be fully tested in local environments. Clear input-output protocols accelerated integration, reducing what traditionally required weeks into a single day of testing. Additionally, built-in workflow visualization and real-time metrics dramatically shortened troubleshooting cycles and streamlined performance optimization, significantly improving both development velocity and system reliability.

Implementing DDP delivered three quantifiable engineering advantages: First, architectural simplification reduced computation units by 50\% (from 20 to 10), dramatically decreasing system complexity and maintenance overhead. Second, this streamlined architecture accelerated development velocity by 40\% compared to the original implementation. Finally, DDP's modular design enabled fully parallel development, allowing team members to work simultaneously on different pipeline components with clear interfaces, eliminating integration bottlenecks that had previously constrained productivity. These improvements collectively transformed both system performance and team efficiency, demonstrating DDP's value beyond mere computational gains.

\begin{center}
\begin{table}

\caption{We present the industry large-scale batch processing results above. D denotes development efficiency, R is runtime performance, and M is system maintainability. mln = million}
\small

\begin{tabular}{ |c| c| c| c| }
\hline
\multicolumn{2}{|c|}{\textbf{Metric and Dimension}}  & \textbf{Native Spark} &  \textbf{DDP} \\
\hline
\multirow{4}{*}{D} & \textbf{\# Computation Units} & 19 & 10 \\ \cline{2-4}
\multirow{4}{*}{} & \textbf{Lines of Code} & 1644 & 930 \\ \cline{2-4}
\multirow{4}{*}{} & \textbf{Development Months} & 4 & 2 \\ \cline{2-4}
\multirow{4}{*}{} & \textbf{Integration Efforts} & 1 week & 1 day \\ \hline
\multirow{2}{*}{R} & \textbf{Scalability Limit} & 1 mln & 500 mln  \\ \cline{2-4}
\multirow{2}{*}{} & \textbf{Latency(1 million)} & 20 hours & 1 hour \\ \hline
M & \textbf{Troubleshooting Efforts} & 1 week & 1 day \\ \hline
\end{tabular}

\label{industry}

\end{table}

\end{center}

\subsection{Web-Scale Language Detection Experiment}
This experiment focuses on detecting and categorizing languages in a massive web document corpus with 2.1 million documents drawn from the Common Crawl Corpus~\footnote{https://paperswithcode.com/dataset/ccnet}. We developed three implementations: one using our DDP framework, and the other two without DDP framework (one using single thread in Python, and one using the distributed Ray system~\cite{moritz2018ray}) to serve as a performance benchmark for comparison.

\subsubsection{Experiment Description}
\begin{figure}[h]
\vspace{1em}
    \centering
    \vspace{-1em}
    \includegraphics[width=8.6cm]{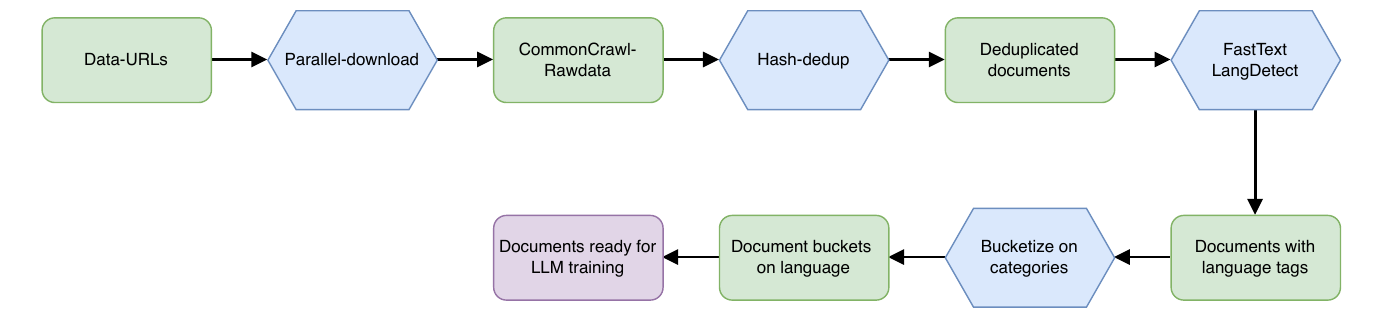}
    \vspace{-1em}
    \caption{We implemented the above data processing stages with DDP in the web-Scale language detection experiment to study the gains over conventional implementations.}
    \label{data_pipe_illustration}
    
\end{figure}

Using the framework, we leveraged the visual design interface to decompose the task into pre-processing subtasks and document subtasks (document deduplication and language detection/partitioning) in \cref{data_pipe_illustration}. For each subtask, we defined the necessary DataDeclare, TransformerDeclare, and MetricDeclare components. The DataDeclare definitions specified the input/output locations, schemas, and file formats for web documents. The TransformerDeclare outlined the transformation logic for deduplication and language detection. The MetricDeclare defined performance metrics such as document counts per language and deduplication rates.

DDP framework automated the execution of subtasks, parallelization of computations, handling data flow management, transformation integration, and metrics collection. It managed the dependencies between the deduplication and language detection processes well, ensuring optimal execution order and resource utilization.

In contrast, the non-framework implementation required manual coding for task orchestration, data handling, and process coordination. This implementation lacked built-in parallelization capabilities and real-time metrics monitoring, making it more challenging to manage and optimize the processing pipeline.

\subsubsection{Experiment Settings}
We use Ray version 2.41.0, Glue version 5.0, Spark version 3.0. Cluster sizes ranges from 1 to 48, and the instance type is G.1X, which is similar to the m4.2xlarge EC2 instance in terms of price and configurations.

\subsubsection{Experiment Analysis}
From a development efficiency perspective, the DDP implementation required 175 lines of code compared to 245 lines for the non-DDP approach. The visual design interface enabled 100\% task parallelism, allowing two developers to work independently on deduplication and language detection. The DDP's modular structure facilitated comprehensive testing, achieving 95\% test coverage, while the non-DDP implementation achieved only 70\% coverage due to its monolithic nature.

\begin{figure}[h]
    \centering
    \vspace{-1em}
    \includegraphics[width=8.6cm]
    {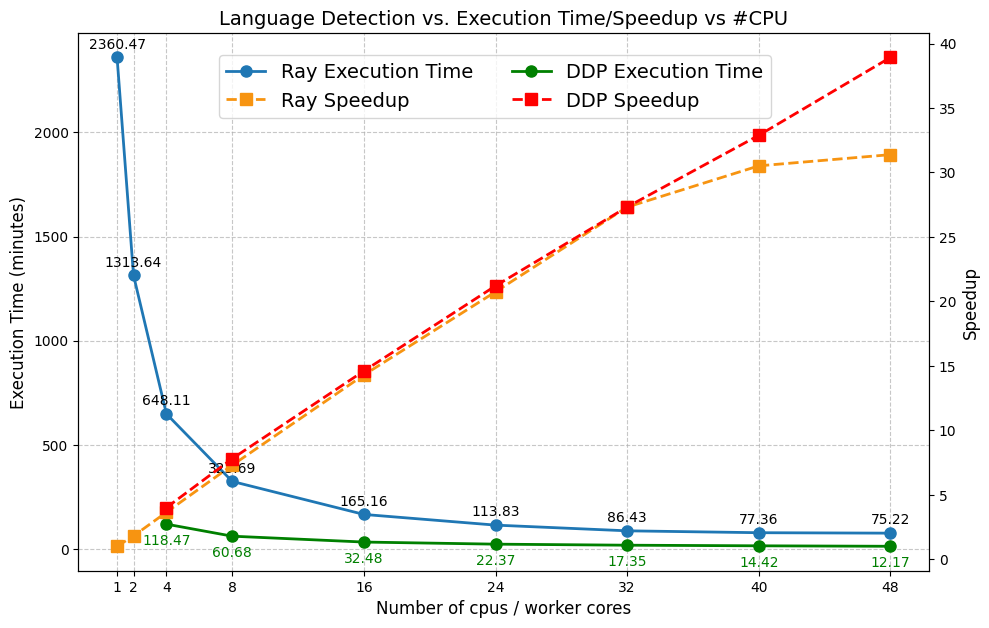}
    \vspace{-1em}
    \caption{Scalability Evaluation over 2.1 M documents from the CC-NET corpus. The smallest number of CPUs for our DDP framework was 4 (single worker instance).}
    
    \label{fig:exp2}
\end{figure}

Runtime performance analysis showed significant improvements with the DDP framework. As shown in Table \ref{web}, using 12 G.1X AWS Glue workers with total 48 vcpu, the framework completed processing in 13 minutes, compared to non-DDP implementations such as 2360 minutes for single thread implementation and 75 minutes for Ray-based implementation both on a single instance with 48 vcpu. Our framework achieved an average CPU utilization of 99\%, demonstrating efficient resource usage through optimized parallel processing.

For system maintenance, The DDP framework provided comprehensive monitoring through built-in metrics publishing. Key metrics included document counts per language, deduplication rates, and processing throughput, all updated at 30-second intervals. Adding new metrics required minimal effort due to the framework's declarative approach. In contrast, The non-DDP implementation did not provide metrics publishing functionality, requiring substantial additional code for monitoring capabilities.

\begin{center}

\begin{table}
\caption{Web-Scale Language Detection Experiment}
    \small
\begin{tabular}{ |c| c| c| c| }
\hline
\textbf{Metric} & \textbf{Python} &  \textbf{DDP} & \textbf{Ray} \\
\hline
\textbf{Lines of Code} & 245 & 175 & 300 \\ \hline
\textbf{Task Parallelism Rate} & 0\% & 100\% & 100\% \\ \hline
\textbf{Execution Time} & 2360 mins & 13 mins & 75 mins  \\  \hline
\textbf{CPU utilization} & 11.9\% & 99\% & 89\%  \\ \hline
\textbf{Number of Cores} & 1 & 48 & 48  \\ \hline
\end{tabular}

\label{web}

\end{table}

\end{center}

The experimental results demonstrate that our DDP framework significantly improved development efficiency, runtime performance, and system maintainability. The visual design interface facilitated parallel development, while the framework's automated handling of data flow and metrics collection reduced complexity and improved monitoring capabilities. The performance gains in both processing time and resource utilization validate the framework's effectiveness for large-scale language detection tasks.

\subsection{Future Work: Hosting LLMs}

We conduct prelimiary experiments on integrating DDP with Llama.cpp to support hosting LLM, each worker instance loaded the LLM model inside the system memory for CPU, or inside GPU. We deploy Qwen2.5-7B-Instruct-f16 model on EMR cluster that consists of 100 c7i.8x instances without GPU. We tested 5000 English to Chinese translation tasks, and the overall latency is 10 hours. Another experiment loaded Qwen2.5-7B-Instruct-f16 model on EMR cluster that consists 6 g6e.8x instances with Nvidia L40S GPU per instance. We tested 5000 English to Chinese translation tasks, and the overall latency is 2 hours. In this distributed system, we treat the model as one single pipe and integrate with other upstream or downstream pipes as a batch processing system. We note that this case study showcases the potential of applying DDP in LLM-related services and leave further development to future work.
\section{Launched Products Using This Architecture}

DDP served as the infrastructure for a list of public ML services in cloud computing, such as: (1) a knowledge graph service that helps organizations match and link records stored across multiple knowledge base, applications, channels and data stores; (2) a self-configurable rule-based data matching service that supports various algorithms (Levenshtein Distance, Cosine Similarity, etc); (3) a data collaboration service that enables privacy-preserved machine learning. These ML services share a common technical challenge: managing the quadratic complexity (O(N²)) that emerges from necessary pairwise computations across N data records. Compared with native Spark, DDP reduced the development effort/time by an average of 40\% while enabling the billion-scale ML inference within hours.

\vspace{-1em}
\section{Conclusion}
\vspace{-0.5em}
This paper presents a novel declarative pipeline architecture that successfully resolves the traditional trade-off between system performance and developer productivity in modern data processing systems. The enterprise case study and the web-scale language detection experiment, demonstrates significant improvements across multiple dimensions.

The modular ``pipe" architecture of our framework improved the development productivity and enterprise collaboration to reduce the development cycle by 50\%. Runtime performance showed marked improvements in scalability and latency reduction. Our innovative ML model integration approach delivered a 10x throughput improvement over microservice-based solutions, while maintaining 99\% CPU utilization. The framework's built-in support for near-real-time metrics publishing and workflow visualization significantly improved system maintainability and troubleshooting with minimal additional code.

These results demonstrate that through thoughtful architecture design, it is possible to achieve both high performance and excellent developer productivity in modern data processing systems. Future work could extend this approach to real-time streaming scenarios and more complex ML model architectures while further exploring automated testing strategies for distributed systems.



\vspace{-0.5em}
\section{Limitations}

The framework proposed in our work demonstrates several key advantages over automated execution-optimization systems such as SystemML~\cite{boehm2016systemml} and Cedar~\cite{zhao2024cedar}. This includes better balance between performance and maintainability thanks to pre-defined data and execution contracts across pipes, more predictable behavior, superior support for parallel development, and more efficient ML model integration. We note that it also has limitations: (1) it may not achieve the same level of automatic optimization as SystemML and offers less flexibility than Cedar's dynamic operator composition; (2) this approach is physically monolithic compared to micro-service based approach, therefore having higher risk of single-point-of failure; (3) this approach benefits more large data with complicated computational logic and large group collaborations, meaning Its advantage in small scale problem can be limited due to its overhead cost. Our system's structure is more rigid at execution-time due to predefined pipe contracts. However, our work presents a more practical solution for enterprise environments where development efficiency and system maintainability are as crucial as runtime performance

In future work, we will extend DDP to GPU-based tasks like large-scale pretraining, and streaming tasks like real-time analytics.

\bibliography{sample}

\nocite{langley00}

\bibliographystyle{mlsys2025}

\end{document}